\documentclass[12pt,preprint]{aastex}

\def\la{\mathrel{\hbox{\rlap{\hbox{\lower4pt\hbox{$\sim$}}}\hbox{$<$}}}}
\def\ga{\mathrel{\hbox{\rlap{\hbox{\lower4pt\hbox{$\sim$}}}\hbox{$>$}}}}

\begin{document}

\title{Comment on astro-ph/0507588 and astro-ph/0508483}

\author{Daniel E. Reichart\altaffilmark{1}}

\altaffiltext{1}{Department of Physics and Astronomy, University of North
Carolina at Chapel Hill, Campus Box 3255, Chapel Hill, NC 27599;
reichart@physics.unc.edu}

Guidorzi has now written two papers (astro-ph/0507588 and astro-ph/0508483, both accepted to MNRAS) on the GRB variability-luminosity correlation in which he finds that expanded samples of $L$ vs.\ $V$ data are not well described by a power law because the scatter of the data around such a model is more than can be accounted for by the data's statistical errors alone (sample variance) -- ``in contrast with the original findings by Reichart et al.\ (2001)'' -- but then proceeds to model these data with a power law anyway and finds significantly shallower $L$ vs.\ $V$ relationships than Reichart et al.\ (2001) found.  However, as Reichart \& Nysewander (2005; astro-ph/0508111) pointed out after Guidorzi's first posting but before his second, Reichart et al.\ (2001) $never$ modeled their $L$ vs.\ $V$ data with a power law.  Instead, they used a power law with a distribution around it to accommodate and measure this sample variance.  Ignoring sample variance in a fit that requires it very easily results in incorrect fitted parameter values due to increased sensitivity to outliers, as well as significantly underestimated uncertainties in these fitted parameter values.  Fitting to Guidorzi's own data, Reichart \& Nysewander (2005) showed that when sample variance is included in the model, $L \sim V^{3.4^{+0.9}_{-0.6}}$ with a sample variance of $\sigma_{\log{V}} = 0.20^{+0.04}_{-0.04}$, which is in excellent agreement with the original finding of Reichart et al.\ (2001) -- $L \sim V^{3.3^{+1.1}_{-0.9}}$ with a sample variance of $\sigma_{\log{V}} = 0.18^{+0.07}_{-0.05}$ -- when the sample was approximately one-third its current size.  Incorrectly assuming the sample variance to be zero, Guidorzi et al.\ (2005; astro-ph/0507588) find that $L \propto V^{1.3^{+0.8}_{-0.4}}$ and Guidorzi (2005; astro-ph/0508483) find that $L \propto V^{0.85^{+0.02}_{-0.02}}$.  The significantly underestimated uncertainty in the latter fitted value is again a telltale sign of this all-too-common mistake in astronomy.

Before posting a third paper, please reread Reichart et al. (2001) and read Reichart \& Nysewander (2005).

\acknowledgements
DER very gratefully acknowledges
support from NSF's MRI, CAREER, PREST, and REU programs, NASA's APRA, Swift
GI
and IDEAS programs, and especially Leonard Goodman and Henry Cox.

\end{document}